\documentclass[english]{svjour3}
\usepackage[T1]{fontenc}
\usepackage[latin9]{inputenc}
\usepackage{float}
\usepackage{url}
\usepackage{amstext}
\usepackage{amssymb}
\usepackage{graphicx}
\usepackage{esint}

\makeatletter
\RequirePackage{fix-cm}

\smartqed  

\makeatother

\usepackage{babel}
\begin{document}

\title{Protecting coherence by reservoir engineering: intense bath disturbance}

\titlerunning{Protecting coherence by intense bath disturbance}

\author{Zixian Zhou \and Zhiguo L\"{u} \and Hang Zheng}

\institute{Zixian Zhou \and Zhiguo L\"{u} \and Hang Zheng \at Key Laboratory
of Artificial Structures and Quantum Control (Ministry of Education),
Department of Physics and Astronomy, Shanghai Jiao Tong University,
Shanghai 200240, People's Republic of China\\
Collaborative Innovation Center of Advanced Microstructures, Nanjing
University, Nanjing 210093, People's Republic of China\\
\email{zzx1313@sjtu.edu.cn}\\
\and Zhiguo L\"{u}\\
\email{zglv@sjtu.edu.cn}\\
\\
Hang Zheng\\
\email{hzheng@sjtu.edu.cn}}

\date{Received: date / Accepted: date}
\maketitle
\begin{abstract}
We put forward a scheme based on reservoir engineering to protect
quantum coherence from leaking to bath, in which we intensely disturb
the Lorentzian bath by $N$ harmonic oscillators. We show that the
intense disturbance changes the spectrum of the bath and reduces the
qubit-bath interaction. Furthermore, we give the exact time evolution
with the Lorentzian spectrum by a master equation, and calculate the
concurrence and survival probability of the qubits to demonstrate
the effect of the intense bath disturbance on the protection of coherence.
Meanwhile, we reveal the dynamic effects of counter-rotating interaction
on the qubits as compared to the results of the rotating wave approximation.

\end{abstract}

\section{Introduction\label{sec:Introduction}}

Quantum superposition and entanglement are fundamental concepts in
quantum mechanics and lead to many interesting results such as Schr\"{o}dinger's
cat \cite{Cat}. They also play an important role in quantum computation
and have a large value of applications in quantum information processing
\cite{Q entg,ESD,Q open sys,ESD_expr}. Actually, it is inevitable
to lose quantum information and entanglement because of the coupling
of a system to a dissipative environment. Various approaches have
been explored to prolong the quantum information, such as the quantum
control \cite{Q control,Q state} and quantum Zeno effect \cite{QZE,prot via QZE,QZE and AQZE}.
The idea of the quantum control is resorting to a series of strong
pulses on the qubits to maintain the quantum information stored in
it. The approach of quantum Zeno effect resorts to a series of projective
measurements on the qubits, and it has been unified with quantum control
\cite{QZE and QC}. Besides the maintenance of the initial coherence,
the steady superposition states can be produced in the existence of
the dissipative environment. This strategy corresponds to the adjustment
of qubit-bath coupling with the help of external laser \cite{rsv eng 1,rsv eng 2,rsv eng 3},
which is known as reservoir engineering.

Till now, the approaches that keep the initial quantum coherence mainly
resort to the operations on the qubits. Then a question arises: can
we engineer the bath to protect the initial coherence? There is a
great advantage of engineering the bath in the coherence protection,
for in keeping the multi-qubit coherence, the strategy of quantum
control requires the operations on every qubit, while engineering
the bath only tackles one common bath. Therefore, it is a more economic
way to protect the quantum entanglement. To realize this effect, we
may borrow the ideas of quantum control and Zeno effect, where the
qubits need to suffer a sufficiently strong coupling to a bunch of
laser beam or detecting apparatus \cite{QZE_Prot} and the initial
coherence is then prolonged. If an intense disturbance is imposed
on the bath, will the quantum information stored in the qubits be
better preserved? We will answer the question in the following.

In this paper, we put forward a scheme based on reservoir engineering
to protect the initial quantum coherence: an intense disturbance to
the bath that couples to two qubits. We describe the disturbance as
$N$ harmonic oscillators which quadratically couple to the bath.
The scheme can be realized by both optical and mechanical ways. For
the optical bath, a bunch of Rydberg atoms can be designed to couple
to the optical cavity \cite{Rydberg}, and the large dipole moment
of the Rydberg atoms provides a sufficient intense coupling. The character
frequencies of the optical cavity and Rydberg atoms are both of GHz.
This designation has been used for the single photon detection, while
the process of detection also reacts to the optical cavity and disturbs
the bath. For the mechanical realization, we design a superconducting
microwave resonator coupling to a mechanical cavity \cite{optmec}.
The microwave resonator has been applied for phonon detection, and
it also causes a disturbance to the mechanical cavity. The character
frequencies of nanomechanical resonator and microwave resonator are
both of 10-100 MHz. And there are many other optomechanical systems
available for the realization listed in Ref. \cite{optmec} with the
character frequencies ranged from kHz to GHz. As showing in the following,
if the bath disturbance is sufficiently strong, the qubit-bath interaction
will be overwhelmed so that the initial quantum entanglement will
be better preserved in the qubits. We will calculate the time evolution
of the qubits to show this effect of coherence protection.

The qubits along with the bath are modeled as the well known spin-boson
model \cite{SBM} in which the counter-rotating (CR) interaction plays
an important role in quantum dynamics such as entanglement sudden
death (ESD) \cite{ESD} and entanglement creation \cite{ESD and C}.
Recent studies have solved the dynamics beyond the rotating-wave approximation
(RWA) \cite{CR 1,CR 2,Lrtz Spc}. In this paper, we choose the Lorentzian
spectrum for the bath because it is corresponding to a damped harmonic
cavity as the scheme requires \cite{prot via QZE,mst eq,diag}, and
we give the exact dynamics of the spin-boson model by a pseudo-mode
master equation. This master equation has been proved and widely applied
in the RWA \cite{mst eq}, and we extend it to the arbitrary form
of the spin-boson interaction with Lorentzian spectrum. Thus, the
effects of the CR terms can be revealed in the comparison to the previous
RWA results.

The paper is organized as follows. In Sect. \ref{sec:Model}, we give
and reduce the Hamiltonian of the model. The influence of the intense
bath disturbance is analyzed. In Sect. \ref{sec:Evolution}, we discuss
the existing condition of dark state and provide the exact solutions
of the concurrence and survival probability. The results are presented
and compared to show the effect of intense bath disturbance and the
dynamical effects of CR terms. In Sect. \ref{sec:Conclusions}, we
make the conclusions.

\section{Model\label{sec:Model}}

\subsection{Hamiltonian}

The model consists of two parts, a two-spin-boson model and the intense
bath disturbance, an intense coupling from oscillators. Its Hamiltonian
in natural unit ($\hbar=c=1$) reads
\begin{equation}
H=H_{SB}+H_{O},\label{eq:H}
\end{equation}
in which the spin-boson Hamiltonian takes the form of 
\begin{equation}
H_{SB}=H_{S}+H_{B}+H_{I},
\end{equation}
\begin{equation}
H_{S}=\frac{\Delta}{2}\left(\sigma_{z}^{\left(1\right)}+\sigma_{z}^{\left(2\right)}\right),
\end{equation}
\begin{equation}
H_{B}=\sum_{k}\omega_{k}b_{k}^{\dagger}b_{k},
\end{equation}
\begin{equation}
H_{I}=g\sum_{j=1,2}\alpha_{j}\sigma_{x}^{\left(j\right)}\sum_{k}\mu_{k}\left(b_{k}+b_{k}^{\dagger}\right),\label{eq:H_I}
\end{equation}
with the qubit frequency $\Delta$, coupling constant $g$ in the
unit of frequency and the normalized real coefficients $\alpha_{1}^{2}+\alpha_{2}^{2}=1$.
Operator $\sigma_{x,z}$ and $b_{k}$ are Pauli matrices and annihilation
operator, respectively. The spectrum is chosen as a standard Lorentzian
type, given by

\begin{equation}
J_{B}\left(\omega\right)=\sum_{k}\mu_{k}^{2}\delta\left(\omega_{k}-\omega\right)=\frac{2\omega_{0}}{\pi}\frac{2\Gamma\omega}{\left(\omega^{2}-\omega_{0}^{2}\right)^{2}+\left(2\Gamma\omega\right)^{2}}\theta\left(\omega\right),\label{eq:J_B}
\end{equation}
with central frequency $\omega_{0}$, decay rate $\Gamma$ and step
function $\theta\left(\omega\right)$. Here $\mu_{k}$ is dimensionless.

\begin{figure}[H]
\includegraphics[width=12cm]{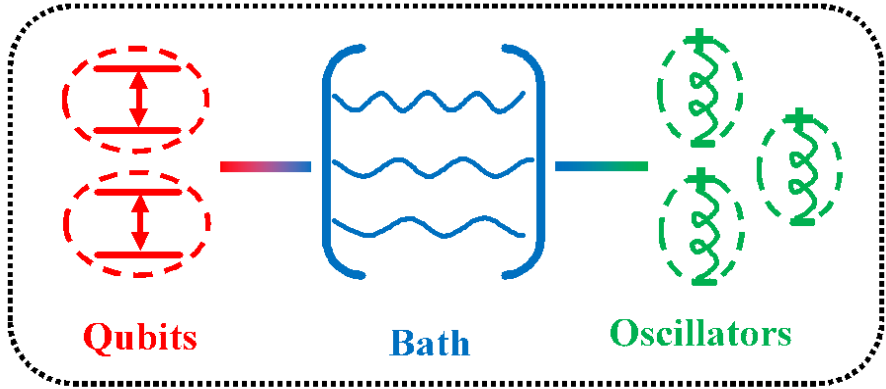}

\protect\caption{\label{fig:sketch}(Color online). Sketch of the optical realization
for the spin-boson model with intense bath disturbance.}
\end{figure}

The disturbance part in Hamiltonian Eq. (\ref{eq:H}) describes $N$
identical oscillators quadratically coupling to the bath, which takes
the form of 
\begin{equation}
H_{O}=\sum_{n=1}^{N}\left[\frac{1}{2m}\left(p_{n}-eA\right)^{2}+\frac{m\Omega^{2}}{2}q_{n}^{2}\right],\label{eq:H_O_opt}
\end{equation}

\begin{equation}
-eA/\sqrt{m}=\sqrt{2W}\sum\limits _{k}\mu_{k}\left(b_{k}+b_{k}^{\dagger}\right),\label{eq:eA}
\end{equation}
with oscillation amplitude $q_{n}$, momentum $p_{n}$, harmonic frequency
$\Omega$, and coupling intensity $\sqrt{\Omega W}$. The increase
of either the parameter $W$ or the number of oscillators $N$ strengthens
the coupling, therefore, we just define $I=NW$ as the total intensity
of the disturbance. The amplitude of cavity field is the same one
that couples to the qubits, and the spectrum in Eq. (\ref{eq:eA})
is the same one in Eq. (\ref{eq:H_I}). In strong coupling regime,
$\left(eA\right)^{2}$ term cannot be ignored because it keeps the
Hamiltonian positive defined. This simplest model can be realized
by an optical cavity detected by Rydberg atoms \cite{Rydberg}, as
sketched in Fig. \ref{fig:sketch}. The present notation $A$ just
denotes electromagnetic potential, and the oscillators describe the
Rydberg atoms as dipoles.

\begin{figure}[H]
\includegraphics[height=8cm]{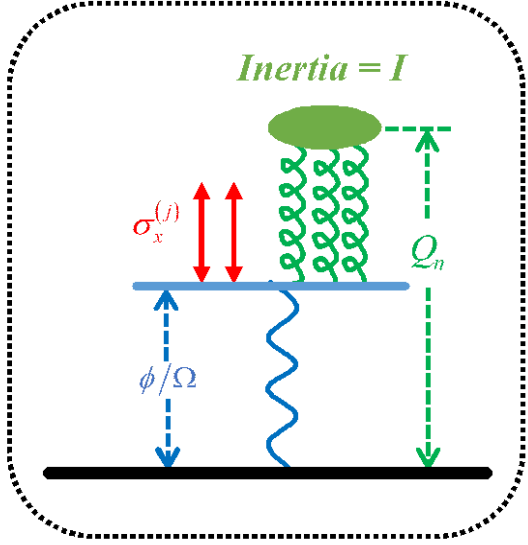}

\protect\caption{\label{fig:illustration}(Color online). Illustration of the mechanism
of reservoir engineering.}
\end{figure}

Besides the optical realization, the model can also be realized by
optomechanical systems \cite{optmec,mec realz}. Ref. \cite{mec realz}
has proposed the scheme that a nanomechanical resonator, coupled capacitively
to an artificial atom, is detected by a superconducting microwave
resonator. In a similar way, we propose that $N$ identical single-mode
microwave resonators compose a parallel circuit and then couple to
a mechanical resonator with two qubits inside. From the present experimental
research \cite{parallel}, we believe that our scheme can be realized
in principle by superconducting quantum circuits. In this case, Hamiltonian
Eq. (\ref{eq:H_O_opt}) is mapped to
\begin{equation}
H_{O}=\sum\limits _{n=1}^{N}\left[\frac{P_{n}^{2}}{2W}+\frac{W\Omega^{2}}{2}\left(Q_{n}-\frac{\phi}{\Omega}\right)^{2}\right],\label{eq:H_O_mec}
\end{equation}

\begin{equation}
\phi=\sqrt{2}\sum\limits _{k}\mu_{k}\left(b_{k}+b_{k}^{\dagger}\right),\label{eq:cvt_field}
\end{equation}
with new coordinates $Q_{n}=-p_{n}/\Omega\sqrt{Wm}$ and momentums
$P_{n}=\Omega\sqrt{Wm}q_{n}$. Here $\phi/\Omega$ in length dimension
($c=1$ used to unify the dimension of time and length) denotes the
phonon field in the nanomechanical resonator, and the $N$ oscillators
describe the $N$ identical single-mode microwave resonators. The
Hamiltonian in Ref. \cite{mec realz} is the RWA form of our Hamiltonian
Eq. (\ref{eq:H_O_mec}). This mechanical system can be simply illustrated
by springs as Fig. \ref{fig:illustration} presents. We can see in
Fig. 2 that all the springs and the qubits are parallel connected
to the phonon field. The disturbance system, parallel springs, generate
a total spring coefficient $NW\Omega^{2}$ and a total mass $NW=I$
which can be seen from Eq. (\ref{eq:H_O_mec}). The large inertia
$I$ from these springs will restrain the oscillation of the phonon
field $\phi/\Omega$ so that the phonon field can hardly receive the
stimulation from the decay of the qubits. It means the intense bath
disturbance attaches a heavy inertia to the bath, then the qubit-bath
interaction is weakened. Consequently, the qubits are prevented from
decay by this mechanism of reservoir engineering.

\subsection{Diagonalization and modified spectrum}

Now the effect of the intense bath disturbance will be investigated
quantitatively. The quadratic terms of the bosonic operators in the
total Hamiltonian can be diagonalized to normal modes as a new reservoir,
namely, \cite{diag}
\begin{equation}
\ensuremath{H_{R}=H_{B}+H_{O}=\sum\limits _{k}\omega_{k}c_{k}^{\dagger}c_{k}.}\label{eq:H_R}
\end{equation}
At the same time, the amplitude of the cavity field Eq. (\ref{eq:cvt_field})
is re-expressed by
\begin{equation}
\phi=\sqrt{2}\sum\limits _{k}\nu_{k}\left(c_{k}+c_{k}^{\dagger}\right)
\end{equation}
with a new spectrum $J_{R}\left(\omega\right)=\sum\limits _{k}\nu_{k}^{2}\delta\left(\omega_{k}-\omega\right)$.
Thus, the total Hamiltonian reproduces the form of the spin-boson
model, which reads
\begin{equation}
H=\frac{\Delta}{2}\sum\limits _{j}\sigma_{z}^{\left(j\right)}+\sum\limits _{k}\omega_{k}c_{k}^{\dagger}c_{k}+g\sum\limits _{j}\alpha_{j}\sigma_{x}^{\left(j\right)}\sum\limits _{k}\nu_{k}\left(c_{k}+c_{k}^{\dagger}\right).
\end{equation}
The concrete expression of $J_{R}\left(\omega\right)$ derived in
Appendix \ref{sec:Appendix A} is given by
\begin{equation}
J_{R}\left(\omega\right)=\frac{2\omega_{0}}{\pi}\frac{2\Gamma\omega\theta\left(\omega\right)}{\left(\omega^{2}-\omega_{0}^{2}-\frac{4I\omega_{0}\omega^{2}}{\omega^{2}-\Omega^{2}}\right)^{2}+\left(2\Gamma\omega\right)^{2}}.\label{eq:J_R}
\end{equation}
Thus, if the disturbance is turned off, we choose $J_{B}\left(\omega\right)$
for the calculation; if it is turned on, we choose $J_{R}\left(\omega\right)$.
It means that the intense bath disturbance just changes the spectrum
of the bath.

\begin{figure}[H]
\includegraphics[width=12cm]{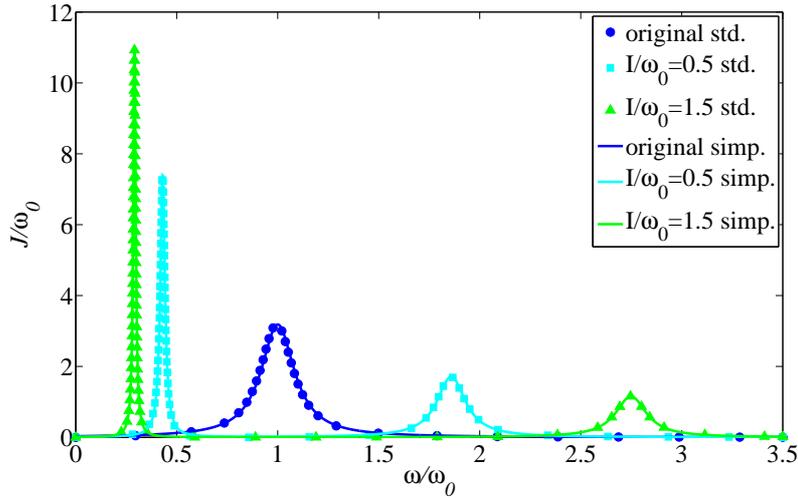}

\protect\caption{\label{fig:spectrum}(Color online). Original spectrum $J_{B}\left(\omega\right)$
with its standard (std.) form Eq. (\ref{eq:J_B}) and simplified (simp.)
form Eq. (\ref{eq:J_B_simp}), and $J_{R}\left(\omega\right)$ with
its standard (std.) form Eq. (\ref{eq:J_R}) and simplified (simp.)
form Eq. (\ref{eq:J_R_simp}), as functions of frequency for $\Gamma/\omega_{0}=0.1$
and $\Omega/\omega_{0}=0.8$.}
\end{figure}

Since the standard Lorentzian spectrum Eq. (\ref{eq:J_B}) is usually
replaced by the simple form \cite{diag}
\begin{equation}
J_{B}\left(\omega\right)=\frac{1}{\pi}\frac{\Gamma}{\left(\omega-\omega_{0}\right)^{2}+\Gamma^{2}},\label{eq:J_B_simp}
\end{equation}
the spectrum Eq. (\ref{eq:J_R}) can be reduced similarly to the summation
of two simplified Lorentzian types (assuming $I$ is sufficient large
to take two peaks apart)
\begin{equation}
J_{R}\left(\omega\right)=\sum_{r=\pm}\frac{\eta_{r}^{2}}{\pi}\frac{\Gamma_{r}}{\left(\omega-\omega_{r}\right)^{2}+\Gamma_{r}^{2}},\label{eq:J_R_simp}
\end{equation}
with the two central frequencies satisfying the equation
\begin{equation}
\omega_{\pm}^{2}-\omega_{0}^{2}-4I\omega_{0}\omega_{\pm}^{2}/\left(\omega_{\pm}^{2}-\Omega^{2}\right)=0,
\end{equation}
the decay rates $\Gamma_{\pm}=\Gamma\left|\omega_{\pm}^{2}-\Omega^{2}\right|/\left|\omega_{+}^{2}-\omega_{-}^{2}\right|$,
and the intensity modification $\eta_{\pm}^{2}=\Gamma_{\pm}\omega_{0}/\Gamma\omega_{\pm}$.
The appearance of the two Lorentzian peaks results from the two normal
modes generated by the boson-oscillator coupling. To confirm the validity
of the simplification, Fig. \ref{fig:spectrum} presents the two expressions
of $J_{B}\left(\omega\right)$, Eq. (\ref{eq:J_B}) and (\ref{eq:J_B_simp}),
as well as the two expressions of $J_{R}\left(\omega\right)$, Eq.(\ref{eq:J_R})
and (\ref{eq:J_R_simp}) for $\Gamma/\omega_{0}=0.1$ and $I/\omega_{0}=0.5,1.5$,
respectively. The dots denote the standard forms and the lines denote
the simplified ones. It is obvious that they agree quite well with
each other, so that it is reasonable to apply the simplified spectrums
in the following discussion.

\begin{figure}[H]
\includegraphics[width=12cm]{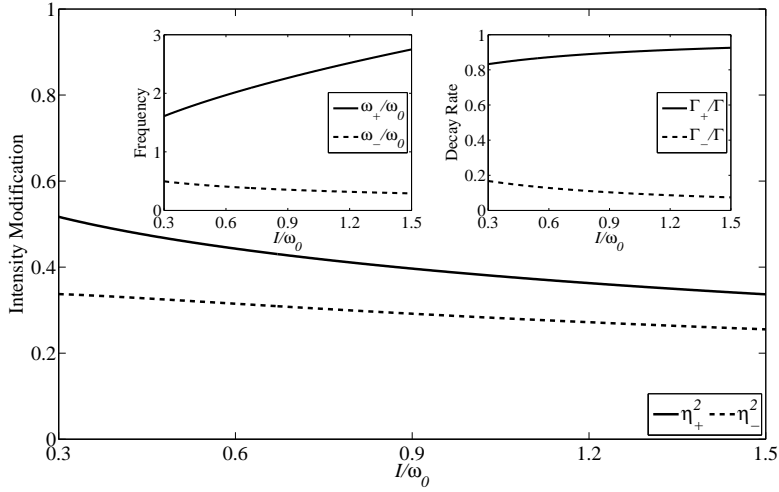}

\protect\caption{\label{fig:parameters}Two central frequencies $\omega_{\pm}$, decay
rates $\Gamma_{\pm}$ and intensity modifications $\eta_{\pm}^{2}$
of the modified spectrum as functions of disturbing intensity $I$
for the same $\Gamma,\Omega$ in Fig. \ref{fig:spectrum}.}
\end{figure}

The intensity modifications $\eta_{\pm}^{2}$, decay rates $\Gamma_{\pm}$,
and central frequencies $\omega_{\pm}$ are plotted in Fig. \ref{fig:parameters}
as functions of the disturbing intensity $I$ for $\Gamma/\omega_{0}=0.1$.
The most important character is the low intensity $\eta_{\pm}$ of
both the peaks, which means the spin-boson coupling is weakened by
the intense bath disturbance. Therefore, the decay of coherence is
suppressed and the quantum information will be better preserved in
the qubits. Besides, the x-coordinate starts from 0.3 because a sufficiently
large $I$ is required to separate the two peaks. As the disturbing
intensity $I$ increases, the two peaks are separated more far away
from each other, with the left peak gradually approaching delta function
and the decay rate of the right peak gradually tending to $\Gamma$. 

\section{Evolution\label{sec:Evolution}}

\subsection{Initial state}

The vacuum state $\left|0_{B}\right\rangle $ of the original bath
$H_{B}$ is engineered by the intense bath disturbance to the new
vacuum state $\left|0_{R}\right\rangle $ corresponding to the new
reservoir $H_{R}$. In this section, we give the evolutions of the
qubits in the off-disturbance Hamiltonian $H_{SB}$ and on-disturbance
Hamiltonian $H$, with the initial product state $\left|\psi\left(0\right)\right\rangle =\left|\psi_{S}\right\rangle \otimes\left|0_{B}\right\rangle $
and $\left|\psi\left(0\right)\right\rangle =\left|\psi_{S}\right\rangle \otimes\left|0_{R}\right\rangle $,
respectively. Here we choose the respective vacuum state $\left|0_{B,R}\right\rangle $
of $H_{B,R}$, for the cavity is usually cooled near to the temperature
of absolutely zero in reality to reduce the decoherence. At first,
we talk about the dark state which is an eigen-state of the total
Hamiltonian so that it does not evolve. From this definition, we can
immediately write down the condition when $\left|\psi\left(0\right)\right\rangle $
is dark:
\begin{equation}
H_{S}\left|\psi_{S}\right\rangle =E\left|\psi_{S}\right\rangle ,\sum\limits _{j}\alpha_{j}\sigma_{x}^{\left(j\right)}\left|\psi_{S}\right\rangle =0.
\end{equation}
The right equation means $\sum\limits _{j}\alpha_{j}\sigma_{x}^{\left(j\right)}$
has eigen-value $0$, so that
\begin{equation}
\det\sum\limits _{j}\alpha_{j}\sigma_{x}^{\left(j\right)}=\left(\alpha_{1}^{2}-\alpha_{2}^{2}\right)^{2}=0,\alpha_{1}=\pm\alpha_{2}.\label{eq:condition}
\end{equation}
This condition causes the destructive interference of the two qubits
so that the state does not evolve. While in the RWA case, there always
exists a dark state despite the value of $\alpha_{j}$, which reads
\begin{equation}
\left|\psi_{-}\right\rangle =\left(-\alpha_{2}\left|10\right\rangle +\alpha_{1}\left|01\right\rangle \right)\otimes\left|0_{B,R}\right\rangle .
\end{equation}
However, it is no longer dark in our system if the condition Eq. (\ref{eq:condition})
is not satisfied. Thus, we just call it subradiant state. Furthermore,
we define the superradiant state \cite{prot via QZE} 
\begin{equation}
\left|\psi_{+}\right\rangle =\left(\alpha_{1}\left|10\right\rangle +\alpha_{2}\left|01\right\rangle \right)\otimes\left|0_{B,R}\right\rangle .
\end{equation}
The initial state is chosen as their linear combination (zero phase
difference for simplicity), which reads 
\begin{equation}
\left|\psi\left(0\right)\right\rangle =\left(\cos\frac{\Theta}{2}\left|10\right\rangle +\sin\frac{\Theta}{2}\left|01\right\rangle \right)\otimes\left|0_{B,R}\right\rangle .
\end{equation}
The evolution of this state for the RWA Hamiltonian has been provided
in Ref. \cite{prot via QZE}. The comparison between the RWA and exact
evolutions will be made to reveal the dynamic effect of the CR terms.
And we will also demonstrate the effect of the intense bath disturbance
by comparing the evolutions in Hamiltonians $H_{SB}$ and $H$.

\subsection{Reduced density operator and concurrence}

The exact evolution of the reduced density operator for spins, which
is derived in Appendix \ref{sec:Appendix B} by taking advantage of
a peculiar property of the Lorentzian spectrum, is given here. When
the intense bath disturbance is turned off, the spectrum has only
one Lorentzian peak. If the initial state takes the form of $\left|\psi\left(0\right)\right\rangle =\left|\psi_{S}\right\rangle \otimes\left|0_{B}\right\rangle $,
the reduced density operator for the system is given by $\rho_{S}\left(t\right)=\mathrm{tr}_{a}\tilde{\rho}\left(t\right)$,
in which $\tilde{\rho}\left(t\right)$ satisfies the pseudo-mode master
equation
\begin{equation}
\frac{d\tilde{\rho}\left(t\right)}{dt}=\frac{1}{i}\left[\tilde{H}_{SB},\tilde{\rho}\left(t\right)\right]-\Gamma\left[a^{\dagger}a\tilde{\rho}\left(t\right)+\tilde{\rho}\left(t\right)a^{\dagger}a-2a\tilde{\rho}\left(t\right)a^{\dagger}\right],\label{eq:master eq}
\end{equation}
with the initial value $\tilde{\rho}\left(0\right)=\left|\psi_{S}\right\rangle \left\langle \psi_{S}\right|\otimes\left|0_{a}\right\rangle \left\langle 0_{a}\right|$.
Here $\mathrm{tr}_{a}$ eliminates the annihilation operator $a$
and $\left|0_{a}\right\rangle $ is its ground state. The replaced
Hamiltonian is given by
\begin{equation}
\tilde{H}_{SB}=H_{S}+\omega_{0}a^{\dagger}a+g\sum\limits _{j}\alpha_{j}\sigma_{x}^{\left(j\right)}\left(a+a^{\dagger}\right),\label{eq:tilde_H_SB}
\end{equation}
which becomes a single-mode version of $H_{SB}$ in which the single-mode
frequency is the central frequency of the Lorentzian peak. Since there
is one single mode, it is easy to do the exact numerical calculation.
Though the master equation Eq. (\ref{eq:master eq}) is the same as
that in Ref. \cite{mst eq} which has been widely used in quantum
optics \cite{Q opti}, it was proved and applied only in the RWA.
Our proof in Appendix \ref{sec:Appendix B} is based on the expansion
of the decay rate $\Gamma$ rather than the coupling constant $g$,
therefore, the proof is independent on the concrete form of spin-boson
interaction. Therefore, this master equation is suitable for a wide
classes of Hamiltonian system, as long as the system-bath coupling
takes the Lorentzian spectrum and the bath is initially prepared in
the vacuum state.

The pseudo-mode master equation Eq. (\ref{eq:master eq}) is straightforwardly
extended to the on-disturbance case when the spectrum splits to two
Lorentzian peaks. If the system is initialized in $\left|\psi_{S}\right\rangle \otimes\left|0_{R}\right\rangle $,
the reduced density operator is given by $\rho_{S}\left(t\right)=\mathrm{tr}_{+}\mathrm{tr}_{-}\tilde{\rho}\left(t\right)$,
in which $\tilde{\rho}\left(t\right)$ satisfies 
\begin{equation}
\frac{d\tilde{\rho}\left(t\right)}{dt}=\frac{1}{i}\left[\tilde{H},\tilde{\rho}\left(t\right)\right]-\sum_{r=\pm}\Gamma_{r}\left[a_{r}^{\dagger}a_{r}\tilde{\rho}\left(t\right)+\tilde{\rho}\left(t\right)a_{r}^{\dagger}a_{r}-2a_{r}\tilde{\rho}\left(t\right)a_{r}^{\dagger}\right],\label{eq:master eq gen}
\end{equation}
with the initial value $\tilde{\rho}\left(0\right)=\left|\psi_{S}\right\rangle \left\langle \psi_{S}\right|\otimes\left|0_{+}\right\rangle \left\langle 0_{+}\right|\otimes\left|0_{-}\right\rangle \left\langle 0_{-}\right|$.
Here the notation $\mathrm{tr}_{\pm}$ traces over the annihilation
operator $a_{\pm}$, respectively, and $\left|0_{\pm}\right\rangle $
is the respective ground state. The replaced Hamiltonian is given
by 
\begin{equation}
\tilde{H}=H_{S}+\sum_{r=\pm}\omega_{r}a_{r}^{\dagger}a_{r}+g\sum\limits _{j}\alpha_{j}\sigma_{x}^{\left(j\right)}\sum_{r=\pm}\eta_{r}\left(a_{r}+a_{r}^{\dagger}\right).
\end{equation}
It is the two-mode version of $H$, in which the two coupling constants
are $g\eta_{\pm}$. As master equation Eq. (\ref{eq:master eq gen})
implies, the time evolution is dominated by the modified coupling
intensities $g\eta_{\pm}$ and decay rates $\mbox{\ensuremath{\Gamma_{\pm}}}$.
Since the qubit-reservoir coupling constants $g\eta_{\pm}$ are reduced
by the intense bath disturbance (see Fig. \ref{fig:parameters}),
quantum information stored in the qubits is prevented from leaking
into the reservoir. Therefore, the intense bath disturbance protects
the quantum coherence.

\begin{figure}[H]
\includegraphics[width=12cm]{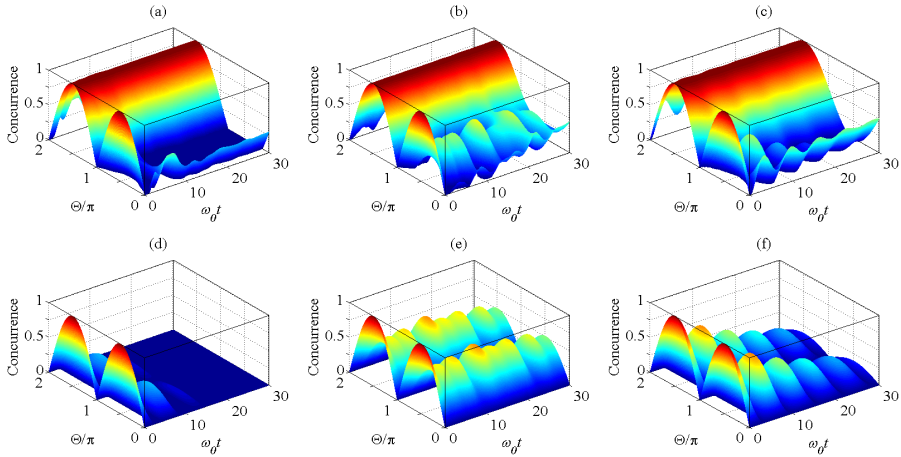}

\protect\caption{\label{fig:concurrence}(Color online). Concurrence as functions of
time $t$ and initial-state parameter $\Theta$ for $\alpha_{1}=\alpha_{2}=1/\sqrt{2}$
in the off-disturbance (a), on-disturbance (b) and the RWA (c) cases,
respectively. Concurrence for $\alpha_{1}=1,\alpha_{2}=0$ in the
off-disturbance (d), on-disturbance (e) and the RWA (f) cases, respectively.
Parameters are set for $\Delta/\omega_{0}=1$, $g/\omega_{0}=0.5$,
$I/\omega_{0}=1.5$, and $\Gamma,\Omega$ are the same ones in Fig.
\ref{fig:parameters}.}
\end{figure}

Till now, we start to investigate the decay and preservation of quantum
entanglement. Concurrence is the magnitude measuring the extent of
entanglement \cite{concur}. Since the density operator takes X-form,
the concurrence has a simple expression which reads \cite{ESD and C}
\begin{equation}
C=2\max\left(0,\left|\rho_{23}\right|-\sqrt{\rho_{11}\rho_{44}},\left|\rho_{14}\right|-\sqrt{\rho_{22}\rho_{33}}\right),
\end{equation}
where $\rho_{ij}$ is the matrix elements of the reduced density operator
$\rho_{S}$. The evolution of concurrence in resonance case ($\Delta=\omega_{0}$)
is presented in Fig. \ref{fig:concurrence} for two sets of parameters,
dark-state-existing $\alpha_{1}=\alpha_{2}=1/\sqrt{2}$ (Figs. \ref{fig:concurrence}
(a)-(c)) and dark-state-disappearing $\alpha_{1}=1,\alpha_{2}=0$
(Figs. \ref{fig:concurrence} (d)-(f)), in which the results of the
off-disturbance, on-disturbance, and the RWA are shown for comparison.

Firstly, we demonstrate the roles of the CR terms on the time evolution
of concurrence by the comparison of the exact off-disturbance result
(Figs. \ref{fig:concurrence} (a) and (d)) with the RWA results (Figs.
\ref{fig:concurrence} (c) and (f)). We find their structures of time
evolution are totally different, for the concurrence with the CR terms
decays very quickly down to zero at and does not revive any more.
This phenomenon is called ESD. While in the RWA case, we find the
ESD does not occur. This is because the RWA interaction forbids the
spin state jumping to $\left|11\right\rangle $, confining the quantum
information in a much smaller subspace, so that it reduces the quantum
entanglement running away. And in this case, the concurrence has a
simpler expression which reads $C_{\text{RWA}}=2\left|\rho_{23}\right|$
\cite{prot via QZE}, so that it reaches zero only at several individual
time points. In other words, the ESD will never occur. Therefore,
the dynamic effect of the CR interaction is revealed: it makes entanglement
decrease more violently and end within finite time.

Secondly, we demonstrate the significant roles of the intense bath
disturbance on the preservation of entanglement by the comparison
of the exact on-disturbance results (Figs. \ref{fig:concurrence}
(b) and (e)) with the off-disturbance results (Figs. \ref{fig:concurrence}
(a) and (d)). In Fig. \ref{fig:concurrence} (a), the concurrence
of the dark state at $\Theta=3\pi/2$ keeps at 1, for the present
set of parameter $\alpha_{1}=\alpha_{2}=1/\sqrt{2}$ satisfies the
dark-state-existing condition Eq. (\ref{eq:condition}), so that the
subradiant state $\left|\psi_{-}\right\rangle $ which corresponds
to $\Theta=3\pi/2$ does not evolve. While for the superradiant state
$\Theta=\pi/2$, its concurrence decays very quickly down to zero
at $\omega_{0}t\approx15$ and undergoes the ESD. The same phenomenon
also happens in Fig. \ref{fig:concurrence} (d), where the ESD phenomenon
appears near at the same time. On the other hand, the intense bath
disturbance successfully keeps the concurrence from falling to zero.
Even at $\omega_{0}t=30$ the concurrence in both Figs. \ref{fig:concurrence}
(b) and (e) maintain at a high level. It means the intense bath disturbance
saves the quantum entanglement. In the previous discussion, we know
the intense bath disturbance attaches a large inertia to the bath,
restraining its excitation, so that the bath can hardly receive the
stimulation from the decay of the qubits. This mechanism modifies
the spectrum of the bath, reducing the intensity of the spectrum.
Thus, the spin-boson interaction is lowered, so that the information
in the qubits is prevented from leaking to the bath. As a result,
the quantum entanglement is well preserved by the intense bath disturbance.

\begin{figure}[H]
\includegraphics[width=12cm]{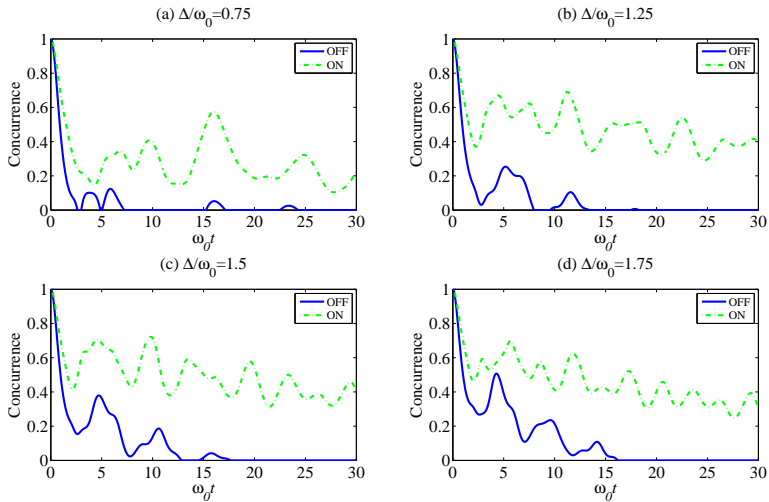}

\protect\caption{\label{fig:ConcurOff}(Color online). Concurrence as functions of
time $t$ for $\Delta/\omega_{0}=0.75$ (a), $\Delta/\omega_{0}=1.25$
(b), $\Delta/\omega_{0}=1.5$ (c) and $\Delta/\omega_{0}=1.75$ (d),
with initial superradiant state $\left|\psi_{+}\right\rangle $ and
$\alpha_{1}=\alpha_{2}=1/\sqrt{2}$. The blue and green lines denote
the off-disturbance (OFF) and on-disturbance (ON) results, respectively.
Parameters $g,I,\Gamma,\Omega$ are the same ones in Fig. \ref{fig:concurrence}.}
\end{figure}

To reveal the effect of the intense bath disturbance in off-resonance
case, we present the evolution of concurrence in Fig. \ref{fig:ConcurOff}
for different values of $\Delta/\omega_{0}$, in which the initial
state is chosen as the superradiant state $\Theta=\pi/2$ when $\alpha_{1}=\alpha_{2}=1/\sqrt{2}$.
The corresponding resonance case has been shown in Fig. \ref{fig:concurrence}
(a). We can see in Figs. \ref{fig:ConcurOff} (a)-(d) that the concurrence
in the off-disturbance case (blue lines) decays quickly down to zero
at $\omega_{0}t\approx15$ in each off-resonance case and undergoes
the ESD phenomenon. On the other hand, the concurrence in the on-disturbance
case (green lines) maintains at a finite level in each case. It means
the detuning between qubit and cavity does not help save the entanglement,
while the effect of the intense bath disturbance saves it. Actually,
the evolution cannot be simply described by an exponential decay with
the rate proportional to $J_{R}\left(\Delta\right)$. We can see from
Eq. (\ref{eq:master eq gen}) that the evolution is mainly dominated
by the heights of the Lorentzian peaks $\eta_{\pm}$ rather than the
central frequencies $\omega_{\pm}$. Since the intense disturbance
reduces the spin-boson interaction $g\eta_{\pm}$, the quantum entanglement
is better preserved in the qubits for both the resonance and off-resonance
cases.

\subsection{Survival probability}

Survival probability is another significant quantity \cite{prot via QZE,QZE and AQZE},
which measures how much the origin quantum information remains, defined
as 
\begin{equation}
P\left(t\right)=\left|\left\langle \psi\left(0\right)|\psi\left(t\right)\right\rangle \right|^{2}.
\end{equation}
It is a simpler dynamic magnitude and independent of the density operator.
Its expression is also derived in Appendix \ref{sec:Appendix B}:
for the off-disturbance case, it is given by 
\begin{equation}
P\left(t\right)=\left|\left\langle \psi_{S}0_{a}\right|\exp\left(-i\tilde{H}_{SB}t-\Gamma a^{\dagger}at\right)\left|\psi_{S}0_{a}\right\rangle \right|^{2};\label{eq:surv prob}
\end{equation}
while for the on-disturbance case, it is straightforwardly extended
to 
\begin{equation}
P\left(t\right)=\left|\left\langle \psi_{S}0_{+}0_{-}\right|\exp\left(-i\tilde{H}t-\sum_{r=\pm}\Gamma_{r}a_{r}^{\dagger}a_{r}t\right)\left|\psi_{S}0_{+}0_{-}\right\rangle \right|^{2}.
\end{equation}

\begin{figure}[H]
\includegraphics[width=12cm]{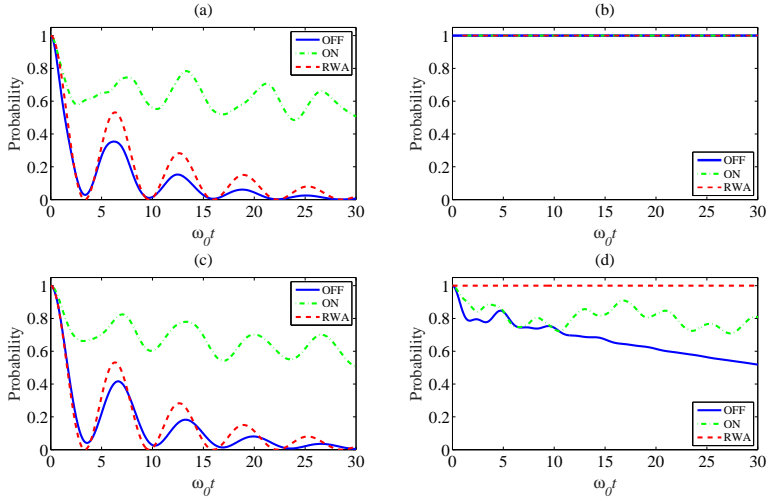}

\protect\caption{\label{fig:probability}(Color online). Survival probability as functions
of time $t$ for $\alpha_{1}=\alpha_{2}=1/\sqrt{2}$ with initial
superradiant state $\left|\psi_{+}\right\rangle $ (a) and subradiant
state $\left|\psi_{-}\right\rangle $ (b). Survival probability for
$\alpha_{1}=1,\alpha_{2}=0$ with initial superradiant state $\left|\psi_{+}\right\rangle $
(c) and subradiant state $\left|\psi_{-}\right\rangle $ (d). The
blue, green and red lines denote the off-disturbance (OFF), on-disturbance
(ON) and RWA results, respectively. Parameters $\Delta,g,I,\Gamma,\Omega$
are the same ones in Fig. \ref{fig:concurrence}.}
\end{figure}

The numerical results of $P\left(t\right)$ in resonance case ($\Delta=\omega_{0}$)
are presented in Fig. \ref{fig:probability} for the same two sets
of $\alpha_{j}$ in Fig. \ref{fig:concurrence} with the corresponding
superradiant states (Figs. \ref{fig:probability} (a) and (c)) and
subradiant states (Figs. \ref{fig:probability} (b) and (d)), in which
the off-disturbance, on-disturbance and RWA cases are shown for comparison.
The probability of the superradiant state is well preserved by the
intense bath disturbance. In Figs. \ref{fig:probability} (a) and
(c), the $P\left(t\right)$ of the off-disturbance and RWA results
decay to almost zero at $\omega_{0}t=30$, while the on-disturbance
results maintain about 50\%. Just like the concurrence, the preservation
of the survival probability also arises from the reduced spin-boson
interaction. In Fig. \ref{fig:probability} (b), all the results have
no time evolution because the dark state condition Eq. (\ref{eq:condition})
is satisfied. However, in Fig. \ref{fig:probability} (d) the off-disturbance
and on-disturbance evolutions decay in comparison with the static
RWA result $P\left(t\right)=1$. These results verify the previous
analysis that dark state only exists in the condition Eq. (\ref{eq:condition}),
which results from the effect of the CR interaction.

\begin{figure}[H]
\includegraphics[width=12cm]{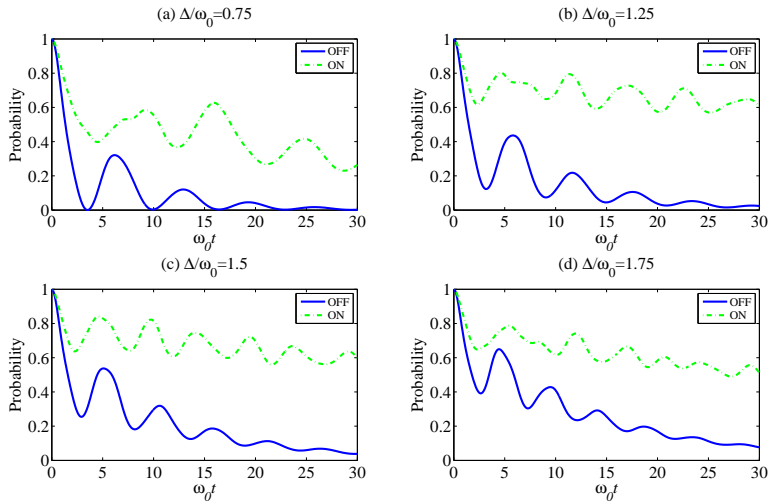}

\protect\caption{\label{fig:ProbOff}(Color online). Survival probability as functions
of time $t$ for $\Delta/\omega_{0}=0.75$ (a), $\Delta/\omega_{0}=1.25$
(b), $\Delta/\omega_{0}=1.5$ (c) and $\Delta/\omega_{0}=1.75$ (d),
with initial superradiant state $\left|\psi_{+}\right\rangle $ and
$\alpha_{1}=\alpha_{2}=1/\sqrt{2}$. The blue and green lines denote
the off-disturbance (OFF) and on-disturbance (ON) results, respectively.
Parameters $g,I,\Gamma,\Omega$ are the same ones in Fig. \ref{fig:ConcurOff}.}
\end{figure}

The evolution of survival probability in off-resonance case is presented
in Fig. \ref{fig:ProbOff} for different values of $\Delta/\omega_{0}$,
in which the initial state is chosen as the superradiant state $\Theta=\pi/2$
when $\alpha_{1}=\alpha_{2}=1/\sqrt{2}$. The corresponding resonance
case has been shown in Fig. \ref{fig:probability} (a). Fig. \ref{fig:ProbOff}
shows that the survival probability in the off-resonance cases is
enhanced at a high level by the intense  bath disturbance. It indicates
that the detuning between the qubits and cavity does not change the
evolution qualitatively, while the intense bath disturbance can suppress
the decay of survival probability. The reason is that the decay process
is mainly dominated by the coupling constants $g\eta_{\pm}$ rather
than free frequencies $\omega_{\pm}$. Since the spin-boson coupling
is reduced by the intense bath disturbance, the quantum information
is better preserved in the qubits, no matter for the resonance or
off-resonance case. Therefore, the detuning effect can be excluded
in the protection of quantum information.

\section{Conclusions\label{sec:Conclusions}}

In summary, we propose a scheme based on reservoir engineering to
protect the initial quantum information and entanglement, in which
the Lorentzian bath is supposed to be intensely disturbed by harmonic
oscillators through a quadratic coupling. The intense bath disturbance
engineers both the vacuum of bath and its excitation, which may intrinsically
change the time evolution of the system. We calculate the time evolution
of the qubits from the initially state $\left|\psi_{S}\right\rangle \otimes\left|0_{R}\right\rangle $,
to compare with that from $\left|\psi_{S}\right\rangle \otimes\left|0_{B}\right\rangle $.
It is revealed qualitatively that the intense bath disturbance increases
the inertia of the bath, restraining the cavity modes in the response
to the stimulation of qubits decaying. Meanwhile, the modified spectrum
of the bath, resulting from the increase of the bath inertia, is obtained
by the diagonalization. And it is found quantitatively that the intensity
of the spectrum is lowered by the intense bath disturbance, which
means the qubit-reservoir interaction is reduced. Therefore, the quantum
information and entanglement stored in the qubits are prevented from
leaking to the environment. The discussed effect has a potential value
of application in quantum computation.

The exact master equation with Lorentzian spectrum is given to solve
the time evolution of the qubits in both the off-disturbance and on-disturbance
cases, in which the multi-modes are converted to one and two pseudo
modes, respectively. The effect of the intense bath disturbance is
manifested in the resonance case, since the on-disturbance evolutions
of both survival probability and concurrence decay much more slowly
than those in the off-disturbance case, which means the quantum information
and entanglement are well preserved by the intense bath disturbance.
Furthermore, the dynamical effects of the CR interaction are also
investigated in the comparison of the off-disturbance evolution and
the RWA results. The CR interaction is found to change the existing
condition of dark state, to accelerate the dissipation, and to cause
the ESD. Finally, the evolution of the concurrence and survival probability
are also shown for the off-resonance cases. The off-disturbance results
still decay rapidly, which excludes the effect of detuning in coherence
protection. And the on-disturbance results decay much more slowly,
which means the intense bath disturbance is also helpful to preserve
the quantum information and entanglement in the off-resonance cases.
Besides, the master equation Eq. (\ref{eq:master eq}) introduced
here will prove useful in treating the properties of certain complicated
models, in particular in the context of time-dependent Hamiltonian.
\begin{acknowledgements}
This work was supported by the National Natural Science Foundation
of China (Grants No. 11174198, No. 11374208, No. 91221201, and No.11474200)
and the National Basic Research Program of China (Grant No. 2011CB922202).
The work was partially supported by the Shanghai Jiao Tong University
SMC-Youth Foundation.
\end{acknowledgements}

\appendix

\section{Derivation of the modified spectrum\label{sec:Appendix A}}

We derive the modified spectrum Eq. (\ref{eq:J_R}) by Green\textquoteright s
function method. Firstly we regard the diagonalized Hamiltonian Eq.
(\ref{eq:H_R}) as total Hamiltonian, which reads
\begin{equation}
H_{R}=H_{A}+H_{B}+H_{C}=\sum\limits _{k}\omega_{k}c_{k}^{\dagger}c_{k},
\end{equation}
\begin{equation}
H_{A}=\sum\limits _{n=1}^{N}\left(\frac{p_{n}^{2}}{2m}+\frac{m\Omega^{2}q_{n}^{2}}{2}\right),
\end{equation}
\begin{equation}
H_{B}=\sum\limits _{k}\omega_{k}b_{k}^{\dagger}b_{k},
\end{equation}
\begin{equation}
H_{C}=P\phi+\frac{I}{2}\phi^{2},\label{eq:H_C}
\end{equation}
with $P=\sqrt{W/m}\sum\limits _{n}p_{n}$ and $\phi=\sqrt{2}\sum\limits _{k}\mu_{k}\left(b_{k}+b_{k}^{\dagger}\right)=\sqrt{2}\sum\limits _{k}\nu_{k}\left(c_{k}+c_{k}^{\dagger}\right)$.
$H_{A}$ is the free Hamiltonian of apparatus. Then we regard $H_{0}=H_{A}+H_{B}$
as the total free Hamiltonian and $H_{C}$ as the interaction part.
Any operator $Q$ in interaction picture (denoted by superscript ``I'')
and Heisenberg picture (denoted by superscript ``H'') has the corresponding
form 
\begin{equation}
Q^{I}\left(t\right)=\exp\left(iH_{0}t\right)Q\exp\left(-iH_{0}t\right),
\end{equation}
\begin{equation}
Q^{H}\left(t\right)=\exp\left(iH_{R}t\right)Q\exp\left(-iH_{R}t\right).
\end{equation}
Next, we define Green\textquoteright s functions 
\begin{equation}
iG_{A}\left(t-t'\right)=\left\langle 0_{A}\right|\hat{T}\left\{ P^{I}\left(t\right)P^{I}\left(t'\right)\right\} \left|0_{A}\right\rangle ,\label{eq:G_A}
\end{equation}
\begin{equation}
iG_{B}\left(t-t'\right)=\left\langle 0_{B}\right|\hat{T}\left\{ \phi^{I}\left(t\right)\phi^{I}\left(t'\right)\right\} \left|0_{B}\right\rangle ,
\end{equation}
\begin{equation}
iG_{R}\left(t-t'\right)=\left\langle 0_{R}\right|\hat{T}\left\{ \phi^{H}\left(t\right)\phi^{H}\left(t'\right)\right\} \left|0_{R}\right\rangle ,
\end{equation}
where $\left|0_{A,B,R}\right\rangle $ is the vacuum state of $H_{A,B,R}$,
respectively. $\hat{T}$ is time-ordering operator. The Green\textquoteright s
function $G_{B,R}\left(\omega\right)$ contains the complete information
of spectrum $J_{B,R}\left(\omega\right)$, respectively. Actually,
the relation between them can be got from their definitions, which
are given by 
\begin{equation}
G_{B,R}\left(\omega\right)=-2\int_{-\infty}^{+\infty}\left(\frac{J_{B,R}\left(z\right)}{z-\omega-i0^{+}}+\frac{J_{B,R}\left(z\right)}{z+\omega-i0^{+}}\right)dz,\label{eq:J to G}
\end{equation}
\begin{equation}
J_{B,R}\left(\omega\right)=-\frac{\theta\left(\omega\right)}{2\pi}\mathrm{Im}G_{B,R}\left(\omega\right),\label{eq:G to J}
\end{equation}
where $G_{A,B,R}\left(\omega\right)=\int_{-\infty}^{+\infty}G_{A,B,R}\left(t\right)\exp\left(i\omega t\right)dt$
is the Fourier transformation of $G_{A,B,R}\left(t\right)$, respectively.
As long as $G_{R}$ is expressed by $G_{A}$ and $G_{B}$, we get
$J_{R}\left(\omega\right)$ immediately. 

Using Gell-Mann-Low theorem, we give \cite{Grn fun} 
\begin{equation}
iG_{R}\left(t-t'\right)=\frac{\sum\limits _{l=0}^{\infty}\frac{1}{i^{l}}\int_{-\infty}^{+\infty}dt_{1}\cdots dt_{l}\left\langle 0_{B}0_{A}\right|\hat{T}\left\{ H_{C}^{I}\left(t_{1}\right)\cdots H_{C}^{I}\left(t_{l}\right)\phi^{I}\left(t\right)\phi^{I}\left(t'\right)\right\} \left|0_{B}0_{A}\right\rangle }{\sum\limits _{l=0}^{\infty}\frac{1}{i^{l}}\int_{-\infty}^{+\infty}dt_{1}\cdots dt_{l}\left\langle 0_{B}0_{A}\right|\hat{T}\left\{ H_{C}^{I}\left(t_{1}\right)\cdots H_{C}^{I}\left(t_{l}\right)\right\} \left|0_{B}0_{A}\right\rangle }.
\end{equation}
Applying Wick\textquoteright s theorem and drawing Feynman diagrams,
we arrive at Dyson equation 
\begin{equation}
G_{R}\left(\omega\right)^{-1}=G_{B}\left(\omega\right)^{-1}-\Sigma\left(\omega\right).\label{eq:Dyson eq}
\end{equation}
Due to the quadratic interaction Eq. (\ref{eq:H_C}), it is easy to
write down the self-energy 
\begin{equation}
\Sigma\left(t-\tau\right)=G_{A}\left(t-\tau\right)+I\delta\left(t-\tau\right),\label{eq:self energy}
\end{equation}
in which $G_{A}$ is got from its definition Eq. (\ref{eq:G_A}),
\begin{equation}
G_{A}\left(\omega\right)=\frac{I\Omega^{2}}{\omega^{2}-\Omega^{2}+i0^{+}}.\label{eq:G_A(w)}
\end{equation}
Inserting the original spectrum Eq. (\ref{eq:J_B}) into Eq. (\ref{eq:J to G}),
we get 
\begin{equation}
G_{B}\left(\omega\right)=\frac{4\omega_{0}}{\omega^{2}-\omega_{0}^{2}+2i\Gamma\left|\omega\right|}.
\end{equation}
Then substituting these results to Eq. (\ref{eq:Dyson eq}) and (\ref{eq:self energy}),
we get the Green\textquoteright s function 
\begin{equation}
G_{R}\left(\omega\right)=\frac{4\omega_{0}}{\omega^{2}-\omega_{0}^{2}-\frac{4I\omega_{0}\omega^{2}}{\omega^{2}-\Omega^{2}}+2i\Gamma\left|\omega\right|}.
\end{equation}
Substituting it back to Eq. (\ref{eq:G to J}), we solve the final
spectrum which is given by 
\begin{equation}
J_{R}\left(\omega\right)=\frac{2\omega_{0}}{\pi}\frac{2\Gamma\omega\theta\left(\omega\right)}{\left(\omega^{2}-\omega_{0}^{2}-\frac{4I\omega_{0}\omega^{2}}{\omega^{2}-\Omega^{2}}\right)^{2}+\left(2\Gamma\omega\right)^{2}}.
\end{equation}
It is noticeable from Eq. (\ref{eq:self energy}) and (\ref{eq:G_A(w)})
that the self-energy of the cavity field is proportional to the coupling
intensity $I$, which means the intense bath disturbance added to
the cavity increases the inertia of the cavity.

\section{Exact reduced density operator and survival probability\label{sec:Appendix B}}

We derive the master equation Eq. (\ref{eq:master eq}) here with
the help of Lorentzian spectrum to give the exact evolution of the
reduced density operator. Actually, Lorentzian spectrum results from
the re-expression of the annihilation operator $a=\sum\limits _{k}\mu_{k}b_{k}$
after the diagonalization \cite{diag}: 
\begin{equation}
H_{B}=\omega_{0}a^{\dagger}a+\int_{-\infty}^{+\infty}\omega d\left(\omega\right)^{\dagger}d\left(\omega\right)d\omega+\left[a\int_{-\infty}^{+\infty}\sqrt{\Gamma/\pi}d\left(\omega\right)^{\dagger}d\omega+h.c.\right]=\sum_{k}\omega_{k}b_{k}^{\dagger}b_{k},
\end{equation}
in which $\left[d\left(\omega\right),d\left(\omega'\right)\right]=\left[d\left(\omega\right)^{\dagger},d\left(\omega'\right)^{\dagger}\right]=0$
and $\left[d\left(\omega\right),d\left(\omega'\right)^{\dagger}\right]=\delta\left(\omega-\omega'\right)$.
$H_{B}$ consists of three parts, a single mode, white noise and their
rotating-wave coupling. Using this expression before the diagonalization,
the spin-boson Hamiltonian becomes 
\begin{equation}
H_{SB}=\tilde{H}_{SB}+\int_{-\infty}^{+\infty}\omega d\left(\omega\right)^{\dagger}d\left(\omega\right)d\omega+a\zeta_{-}+a^{\dagger}\zeta_{+},
\end{equation}
with $\tilde{H}_{SB}$ defined in Eq. (\ref{eq:tilde_H_SB}), dissipative
field operators $\zeta_{+}=\sqrt{\Gamma/\pi}\int_{-\infty}^{+\infty}d\left(\omega\right)d\omega$
and $\zeta_{-}=\zeta_{+}^{\dagger}$. Regarding $V=a\zeta_{-}+a^{\dagger}\zeta_{+}$
as the interaction part and turning to interaction picture, we get
the correlation function 
\begin{equation}
\left[\zeta_{+}^{I}\left(t\right),\zeta_{-}^{I}\left(t'\right)\right]=2\Gamma\delta\left(t-t'\right).\label{eq:correlation}
\end{equation}
This instant correlation leads to Markovian dynamics, as will show
in the following.

The density operator in interaction picture $\rho^{I}\left(t\right)$
satisfies the well-known master equation \cite{CR 1} 
\begin{equation}
\frac{d\rho^{I}\left(t\right)}{dt}=\frac{1}{i}\left[V^{I}\left(t\right),\rho\left(0\right)\right]-\int_{0}^{t}dt'\left[V^{I}\left(t\right),\left[V^{I}\left(t'\right),\rho^{I}\left(t'\right)\right]\right].
\end{equation}
Since $\left|0_{B}\right\rangle =\left|0_{a}\right\rangle \otimes\left|0_{D}\right\rangle $
($\left|0_{D}\right\rangle $ is the vacuum state $\zeta_{+}\left|0_{D}\right\rangle =0$),
the initial density operator is given by 
\begin{equation}
\rho\left(0\right)=\left|\psi_{S}\right\rangle \left\langle \psi_{S}\right|\otimes\left|0_{a}\right\rangle \left\langle 0_{a}\right|\otimes\left|0_{D}\right\rangle \left\langle 0_{D}\right|.
\end{equation}
After taking trace over the dissipative field $\mathrm{tr}_{D}$,
we get 
\begin{equation}
\frac{d\tilde{\rho}^{I}\left(t\right)}{dt}=-\int_{0}^{t}dt'\mathrm{tr}{}_{D}\left[V^{I}\left(t\right)V^{I}\left(t'\right)\rho^{I}\left(t'\right)-V^{I}\left(t\right)\rho^{I}\left(t'\right)V^{I}\left(t'\right)\right]-h.c.,\label{eq:temp eq}
\end{equation}
in which $\tilde{\rho}^{I}=\mathrm{tr}_{D}\rho^{I}$. Considering
the instant correlation Eq. (\ref{eq:correlation}), for $t>t'$ one
easily gets 
\begin{equation}
\left[\zeta_{+}^{I}\left(t\right),U\left(t',0\right)\right]=0,\label{eq:commutation}
\end{equation}
in which $U\left(t',0\right)$ is the evolution operator 
\begin{equation}
U\left(t',0\right)=\sum\limits _{l=0}^{\infty}\frac{1}{i^{l}}\int_{0}^{t'}dt_{1}V^{I}\left(t_{1}\right)\int_{0}^{t_{1}}dt_{2}V^{I}\left(t_{2}\right)\cdots\int_{0}^{t_{l-1}}dt_{l}V^{I}\left(t_{l}\right).
\end{equation}
Eq. (\ref{eq:commutation}) means the dissipative field operator can
pass through the evolution operator so that it directly operates on
the initial vacuum state, giving zero, namely, 
\begin{equation}
\zeta_{+}^{I}\left(t\right)\rho^{I}\left(t'\right)=\rho^{I}\left(t'\right)\zeta_{-}^{I}\left(t\right)=0.\label{eq:zeros}
\end{equation}
For the case $t=t'$, $\zeta_{+}^{I}\left(t\right)\rho^{I}\left(t'\right)$
and $\rho^{I}\left(t'\right)\zeta_{-}^{I}\left(t\right)$ are finite
and negligible in integral. Applying the special property Eq. (\ref{eq:zeros}),
we simplify the following terms in Eq. (\ref{eq:temp eq}) 
\begin{equation}
\mathrm{tr}_{D}V^{I}\left(t\right)V^{I}\left(t'\right)\rho^{I}\left(t'\right)=2\Gamma\delta\left(t-t'\right)a^{I}\left(t\right)^{\dagger}a^{I}\left(t'\right)\tilde{\rho}^{I}\left(t'\right),
\end{equation}
\begin{equation}
\mathrm{tr}_{D}V^{I}\left(t\right)\rho^{I}\left(t'\right)V^{I}\left(t'\right)=2\Gamma\delta\left(t-t'\right)a^{I}\left(t\right)\tilde{\rho}^{I}\left(t'\right)a^{I}\left(t'\right)^{\dagger},
\end{equation}
so that Eq. (\ref{eq:temp eq}) becomes a memoryless equation (Markovian
process) which reads 
\begin{equation}
\frac{d\tilde{\rho}^{I}\left(t\right)}{dt}=-\Gamma\left[a^{I}\left(t\right)^{\dagger}a^{I}\left(t\right)\tilde{\rho}^{I}\left(t\right)-a^{I}\left(t\right)\tilde{\rho}^{I}\left(t\right)a^{I}\left(t\right)^{\dagger}\right]-h.c..
\end{equation}
Coming back to the Schr\"{o}dinger\textquoteright s picture, the
master equation becomes 
\begin{equation}
\frac{d\tilde{\rho}\left(t\right)}{dt}=\frac{1}{i}\left[\tilde{H}_{SB},\tilde{\rho}\left(t\right)\right]-\Gamma\left[a^{\dagger}a\tilde{\rho}\left(t\right)+\tilde{\rho}\left(t\right)a^{\dagger}a-2a\tilde{\rho}\left(t\right)a^{\dagger}\right].\label{eq:final eq}
\end{equation}
The reduced density operator is given by $\rho_{S}\left(t\right)=\mathrm{tr}_{a}\tilde{\rho}\left(t\right)$. 

Then we yield the evolution of the survival probability Eq. (\ref{eq:surv prob})
here. We define $p\left(t\right)=\left\langle 0_{D}\right|\rho\left(t\right)\left|0_{D}\right\rangle $
and repeat the deduction from Eq. (\ref{eq:temp eq}) to (\ref{eq:final eq})
with the replacement $\mathrm{tr}_{D}\to\left\langle 0_{D}\right|\cdots\left|0_{D}\right\rangle $,
obtaining a similar result 
\begin{equation}
\frac{dp\left(t\right)}{dt}=\frac{1}{i}\left[\tilde{H}_{SB},p\left(t\right)\right]-\Gamma\left[a^{\dagger}ap\left(t\right)+p\left(t\right)a^{\dagger}a\right].
\end{equation}
Its analytical solution is 
\begin{equation}
p\left(t\right)=\exp\left(i\tilde{H}_{SB}t-\Gamma a^{\dagger}at\right)p\left(0\right)\exp\left(-i\tilde{H}_{SB}t-\Gamma a^{\dagger}at\right).
\end{equation}
Finally the survival probability is given by 
\begin{equation}
P\left(t\right)=\left\langle \psi_{S}0_{a}\right|p\left(t\right)\left|\psi_{S}0_{a}\right\rangle =\left|\left\langle \psi_{S}0_{a}\right|\exp\left(-i\tilde{H}_{SB}t-\Gamma a^{\dagger}at\right)\left|\psi_{S}0_{a}\right\rangle \right|^{2}.
\end{equation}

\end{document}